\documentclass[prb,twocolumn,longbibliography,superscriptaddress,floatfix]{revtex4-2}
\usepackage{color}
\usepackage{amsmath}
\usepackage[english]{babel}
\usepackage{amssymb}
\usepackage{graphicx}
\usepackage{multirow}
\newcommand{\colmod}[1]{\textcolor{black}{#1}}

\begin{document}
\title{Can Silica Nanoparticles Improve Lithium Transport in Polymer Electrolytes?}
\author{Jo\"el Martín Dalmas}
\affiliation{Université Grenoble Alpes, CEA, LITEN, 17 rue des Martyrs, 38054 Grenoble, France}
\author{Ambroise van Roekeghem}
\affiliation{Université Grenoble Alpes, CEA, LITEN, 17 rue des Martyrs, 38054 Grenoble, France}
\author{Natalio Mingo}
\affiliation{Université Grenoble Alpes, CEA, LITEN, 17 rue des Martyrs, 38054 Grenoble, France}
\author{Stefano Mossa$^*$}
\affiliation{Université Grenoble Alpes, CEA, IRIG-MEM-LSim, 38054 Grenoble, France}
\email{stefano.mossa@cea.fr}
\date{\today}
\begin{abstract}
\noindent{\bf ABSTRACT:} The question of whether silica nanoparticles can enhance the ionic conductivity of a polymer electrolyte above its crystallization temperature has remained unclear for the two decades following the first experiments on these systems. We use Molecular Dynamics simulations to decipher the atomic scale mechanisms affecting the properties of LiTFSI-poly(ethylene oxide) electrolytes upon the addition of silica nanoparticles. At any ionic concentration, adding nanoparticles significantly decreases the conductivity. Most of this reduction can be simply accounted for by the diffusion equation, resulting from the fact that the space occupied by the nanoparticles is made inactive and unable to sustain ionic diffusion. We identify two distinct regimes, above and below a concentration threshold, corresponding to very different ionic distributions and coordination features of the various species. The lack of conductivity enhancement observed in the simulations supports the conclusions of some recent measurements, and disagrees with the earliest experimental reports on hybrid silica/polyethylene-oxide electrolytes.
\end{abstract}
\maketitle
\section{Introduction}
\label{sect:introduction}
Developing safer and more efficient batteries has become a top research priority in the quest to decrease fossil fuel use \cite{chu2012opportunities,cabana2010beyond,tarascon2001issues}. A particularly promising direction is the use of lithium-metal anodes, which have a theoretically much higher energy density storage capacity than the conventionally used graphite ones \cite{lu2013review,scrosati2015advances}. However, the lithium-metal anode presents several problems when combined with conventional liquid electrolytes (LE) \cite{lu2013review}: LEs are unable to suppress Li plating and the formation of dendrites, which produces internal short circuits and reduces the efficiency \cite{ghassemi2011real,aurbach2002short,zhang2018advances,steiger2014mechanisms}. LEs are also undesirable because of their high flammability and the risk of leakage \cite{armand2008building,goodenough2010challenges,chu2017path}. In contrast, solid-state electrolytes (SSEs) can avoid such problems~\cite{goodenough2010challenges,murata2000overview}. The two main categories of solid electrolytes are solid polymer electrolytes (SPEs) and ceramic electrolytes (CEs). CEs avoid the safety issues coming from the LEs, and present a high ionic conductivity at room temperature, but their brittle nature makes them hard to process and prone to contact problems with the electrodes \cite{li2015solid} In contrast, SPEs are flexible, easy to manufacture, cheap, and present a better contact with the electrodes \cite{macfarlane1999lithium,kamaya2011lithium}.

The main drawback of SPEs is that their ionic conductivity is much lower than the other available options at room temperature. As an example, PEO-based SPEs have a conductivity of about 10$^{-7}$ to 10$^{-5}$ S/m at room temperature \cite{liang2019recent,maurel2020poly}, while the most commonly used CEs such as NASICON-type LATP (Li$_{1+x}$Al$_{x}$Ti$_{2-x}$(PO$_{4}$)$_{3}$), and LAGP (Li$_{1+x}$Al$_{x}$Ge$_{2-x}$(PO$_{4}$)$_{3}$) \cite{morimoto2013preparation,safanama2016structural}, garnet-type CEs like LLZO (Li$_{7}$La$_{3}$Zr$_{2}$O$_{12}$) \cite{murugan2007fast}, and perovskite-type CEs as LLTO (Li$_{3x}$La$_{(2/3)-x}$TiO$_{3}$) \cite{kwon2017enhanced}, have an ionic conductivity of about 10$^{-5}$ to 10$^{-3}$ S/m. This is still low in comparison with the LEs ionic conductivities, which can reach up to 1~S/m \cite{dahbi2011comparative}.

Hybrid solid electrolytes (HSE) offer a solution to the limitations of both solid polymer electrolytes (SPEs) and ceramic electrolytes (CEs). By combining both materials, HSEs can potentially provide the advantages of each while overcoming their individual drawbacks \cite{liang2019recent,zaman2019visualizing,croce1998nanocomposite,scrosati2001new}. HSEs show great potential as electrolyte candidates due to their ability to combine the mechanical properties and processability of polymers with the enhanced ionic conductivity provided by ceramic fillers. Although research on these hybrid materials began two decades ago \cite{croce1998nanocomposite}, it remains a subject of debate today due to the inconsistent and contradictory results found in the literature.

On the experimental side, various publications support the hypothesis that HSEs enhance ionic conductivity, with both active \cite{zaman2019visualizing} and passive fillers \cite{krawiec1995polymer,jayathilaka2002effect,dissanayake2003effect,croce1998nanocomposite,chung2001enhancement,scrosati2000impedance,croce1999physical,capuano1991composite,appetecchi2000transport,croce2001role} showing promising results. Regarding passive fillers, most of the studies reporting conductivity enhancement date back by two decades. Two main complementary hypotheses have been proposed to explain this behavior. It is well-established that adding ceramic nanoparticles to semi-crystalline PEO decreases the degree of crystallinity of the system \cite{boaretto2021lithium,matsuo1995ionic,ketabi2013effect,zewde2013enhanced}. The amorphous phase resulting from this effect favors ion transport through the bulk, enhancing the global conductivity. This idea has received broad acceptance in the literature. However, the nanoparticles seemed to enhance the conductivity of the polymers even above the melting temperature ($T_m$) \cite{croce2001role}. This first hypothesis therefore does not fully explain the effect of the nanoparticles observed in~\cite{zaman2019visualizing,croce1998nanocomposite}.

The complementary hypothesis is that the addition of nanoparticles leads to the formation of an interphase between the bulk SPE and the nanoparticle surface  \cite{zaman2019visualizing,boaretto2021lithium,han2020recent}. In this region, the PEO matrix is reorganized, creating free spaces that promote ion mobility and leading to the formation of a charge-space region due to the contact with the ceramic surface, further enhancing the ionic conductivity in this surfacial region.

Some recent studies, however, have contradicted this hypothesis, highlighting a negative effect of the nanoparticles on the dynamics of the electrolyte \cite{isaac2022dense,liu2004situ,tekell2023ionic,mogurampelly2015effect,mogurampelly2016influence,mogurampelly2016influence2,fullerton2011influence}, and creating a debate about the existence and the origin of this enhancement of the conductivity. Humidity dependent water uptake has also been found to play an important and non-trivial role in the change of conductivity, both with and without nanoparticle addition \cite{fullerton2011influence}. 

In recent years, computational methods have proven to be a valuable tool for gaining insight into the structural and dynamic properties of polymeric systems \cite{choo2020diffusion}. Some Molecular Dynamics (MD) simulations have been performed on bulk PEO-LiTFSI , but did not consider the effect of the addition of fillers. Instead, they focused on studying the dynamics of both ions and polymers, as well as the local coordination features \cite{diddens2010understanding,brooks2018atomistic}. Despite the substantial literature on experimental research in this field, computational studies are scarce, probably due to the extremely complex nature of HSEs. Although some computational studies have reported an increase in the conductivity \cite{wang2022lithium} when adding nanoparticles, others  have reported the opposite effect \cite{hanson2013mechanisms}. This abundance of conflicting reports about HSEs shows that the literature about this topic is still unclear and inconsistent.
\begin{figure}[t]
\centering
\includegraphics[width=0.48\textwidth]{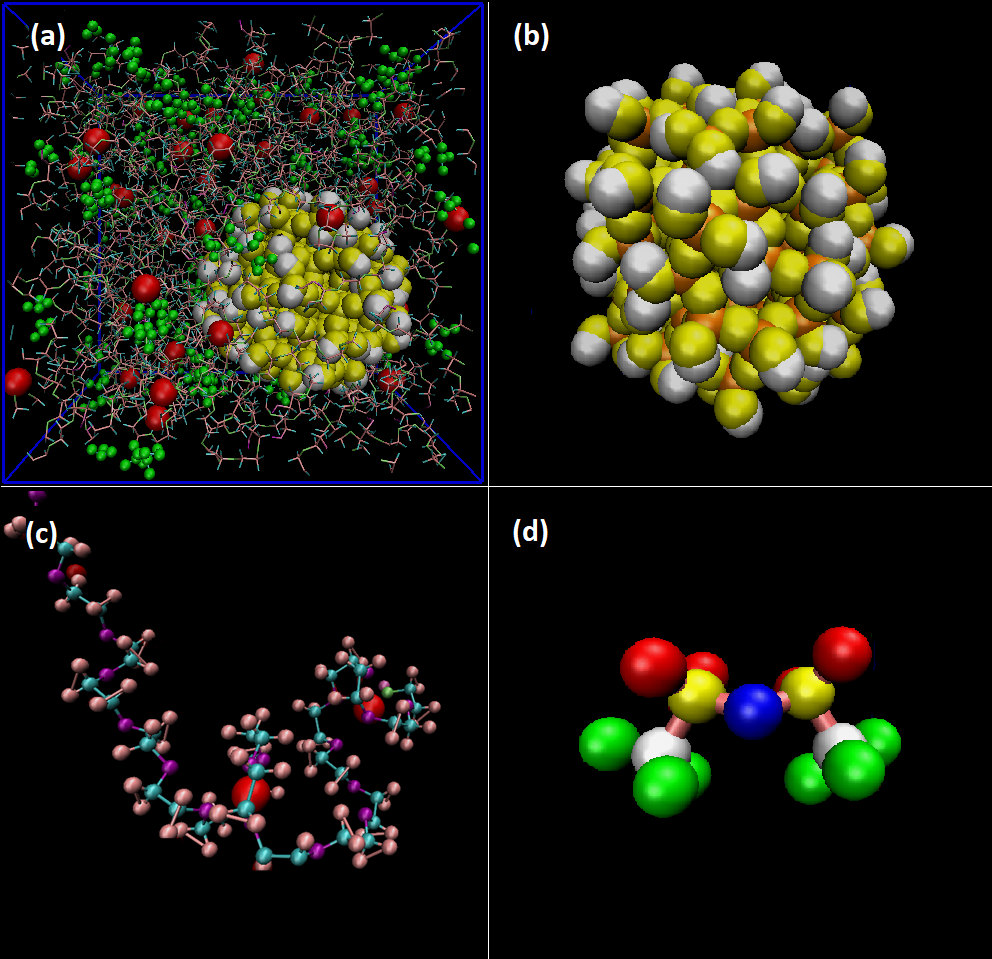}
\caption{Typical views of the investigated systems visualized by VMD~\cite{humphrey1996vmd}. (a) Snapshot of the full system comprising all the elements: lithium, TFSI, SiO$_{2}$ nanoparticle, and PEO. Each element is separately shown in the following panels; (b) SiO$_{2}$ with hydrogen (white), oxygen (yellow) and silicon (orange); (c) Snapshot of the PEO, with the oxygen atoms (purple) coordinating with the lithium ions (red); (d) The TFSI anion with nitrogen (blue), sulfur (yellow), oxygen (red), carbon (white) and fluorine (green) atoms, respectively. The force field employed in the MD simulations and the tools used to initialize the different investigated systems and perform the simulations are discussed in the main text.}
\label{fig:VMD}
\end{figure}

Here, we use Molecular Dynamics (MD) simulations to investigate the properties of PEO-matrix HSEs with SiO$_2$ as a passive filler, across a range of ceramic and ionic concentrations, geometries, and temperatures. By running MD simulations at temperatures above the $T_m$ of PEO we ensure that any possible effect originating from the polymer's crystallinity is eliminated. PEO-based SPEs are most widely studied due to their low glass transition temperature ($T_g$), high solvation properties and excellent interfacial compatibility with electrodes \cite{gadjourova2001ionic,zaghib1999electrochemical,zaghib2004advanced,macglashan1999structure,daigle2015lithium,hovington2015new}. We have chosen SiO$_{2}$ nanoparticles as the filler, due to its low cost, ease of manufacture, high stability, and acidic surface termination \cite{croce1998nanocomposite,liu2004situ,shin2004comparison}. We have selected lithium bis(trifluoromethanesulfonyl)imide (LiTFSI) as the salt, because of its widespread use with PEO in SPEs, high chemical stability, and high dissociation in PEO \cite{lee2019free,falco2019uv,he2019polyethylene,zhang2019composite,li2020progress,diddens2014simulation}.

Although prior computational studies have investigated HSEs~\cite{wang2022lithium,hanson2013mechanisms,li2014atomistic}, to the best of our knowledge none has focused on the use of SiO$_2$ as a filler. Furthermore, previous works have mainly focused on the transport pathways of lithium, given its close interaction with PEO, whereas we provide a broader understanding of the system, studying the dynamical and structural properties for a wide range of ionic and ceramic concentrations, and relating the individual conductivities and the transference number to the coordination number of the different moieties. This work also elucidates the structural effects of adding nanoparticles, particularly the interaction between the SPE and the surface of the ceramic nanoparticles. Our simulations provide atomic-level insight into the structure and dynamics of HSEs, shedding light on the behavior of the intermediate phase between the SPE and the nanoparticle surface. These findings provide important fundamental understanding on the behavior of HSEs, which can contribute to the design of high-performance solid electrolytes for advanced energy storage systems.
\section{Computational Methods}
\label{sect:computational}
\subsection{System components and the force field}
\label{subsect:force field}
We have considered a model hybrid electrolyte consisting  of PEO, TFSI, Li, and SiO$_2$ nanoparticles. Snapshots of all elements are shown in Fig.~\ref{fig:VMD}. The PEO polymer chain consists of 20 ethylene oxide (EO) units and two hydroxyl terminations, for a total of $n_{PEO}=$ 143 atoms. \colmod{(We note that the PEO entanglement length is about $N_e=46$~\cite{tsamopoulos2019shear}.)}. The spherical SiO$_2$ nano-particle has a radius $R_{SiO_2}=$~12~ \AA, and comprises $n_{SiO_2}=$~447 atoms. \colmod{The cation, Li$^+$ has a charge $Q_+=\gamma e$ with $\gamma=0.8$, as customary in non-polarizable force fields}, while the anion, TFSI$^-$, is formed by $n_{TFSI}=15$ atoms and has a total charge $Q_-=- Q_+$. The number of the different components can be tuned in order to explore different relative concentration conditions. In all simulations we have kept constant the number of PEO chains, $N_{PEO}$=100, and considered a large range of conditions varying both the salt content per polymer chain, $\lambda$ (Li:PEO), and the silica nanoparticles volume fraction, $\Phi_{\text{SiO$_2$}}=V_{SiO_2}/V$, with $V_{SiO_2}\simeq$~7238~\AA$^3$ and $V$ the volume of the simulation box.

We have employed the OPLS all-atom force field (OPLS-AA-FF), an empirical, non-polarizable model originally developed for organic molecules and peptides~\cite{jorgensen1996development}. However, due to its accuracy and efficiency, this classical force field has been widely adopted in many areas of polymer science \cite{fang2021revised}, ranging from battery applications to pharmacological research \cite{tzanov2014accurately,robertson2015improved}. The non-bonded interaction pair potential between two atoms $i$ and $j$ of types $\alpha$ and $\beta$, at positions $\vec{r}_i$ and $\vec{r}_j$, is the standard sum of a Lennard Jones and a Coulomb potentials,
\begin{eqnarray}
\label{eq:unbounded}
V_{\text{pair}}(r)&=&V_{LJ}(r)+V_{C}(r)\\
&=& 4 \epsilon_{\alpha\beta} \left[\left(\frac{\sigma_{\alpha\beta}}{r} \right) ^{12}- \left(\frac{\sigma_{\alpha\beta}}{r}\right)^{6}\right] + \frac{C q_i^\alpha q_j^\beta}{\epsilon_o r},\nonumber
\end{eqnarray}
where $r=|\vec{r}_i-\vec{r}_j|$ is the distance, $q_i^\alpha$ and $q_j^\beta$ the charges, and $\epsilon_o$ is the vacuum permittivity. (C is an energy conversion constant.) The employed FF parameters are indicated in Tab.~S1.
We applied the Lorentz-Berthelot mixing rules for the LJ coefficients, $\epsilon_{\alpha\beta}=\sqrt{\epsilon_{\alpha}\epsilon_{\beta}}$ and  $\sigma_{\alpha\beta}=\left(\sigma_{\alpha}+\sigma_{\beta}\right)/2$. We have truncated the terms in Eq.~(\ref{eq:unbounded}) at a cutoff distance $r_c=$10~\AA, while the long-range part of the Coulomb interactions have been computed with a particle-particle particle-mesh solver(PPPM)~\cite{hockney1989particle}, with a relative RMS error in the per-atom forces of 10$^{-4}$. 

Atoms within macro-molecules interact with neighbors separated by up to three bonds via an intramolecular potential, $V_{\text{intra}}=V^b+V^a+V^d$, sum of a stretching, bending and dihedral torsion terms,
\begin{eqnarray}
\label{eq:bonded}
V^b_{ij}&=&K|\vec{r}_{ij}-\vec{r}_{0}|^{2}\nonumber\\ 
V^a_{ijk}&=& K(\theta_{ijk}-\theta_{0})^{2}\\
V^d_{ijkl}&=&\frac{1}{2}K_{1}\left[1+cos(\phi_{ijkl})\right]+\frac{1}{2}K_{2}\left[1-cos(2\phi_{ijkl})\right]\nonumber\\ 
&+& \frac{1}{2}K_{3}\left[1+cos(3\phi_{ijkl})\right]+\frac{1}{2}K_{4}\left[1-cos(4\phi_{ijkl})\right].\nonumber
\end{eqnarray}
The values of the parameters in Eq.~(\ref{eq:bonded}) are reported in the Tabs.~S2 and S3.
We used the standard method of excluding or reducing the pair interactions of Eq.~(\ref{eq:unbounded}) by setting weighting coefficients $(w^b, w^a, w^d)=(0, 0, 1/2)$ for atoms involved in Eq.~(\ref{eq:bonded}), while $w=$~1 for further atoms.
\subsection{Simulation procedure}
\label{subsect: simulation procedure}
Molecular dynamics simulations were carried out using the LAMMPS~\cite{thompson2022lammps} simulation package, applying periodic boundary conditions in the three dimensions. All simulations were  conducted in the (NPT)-ensemble with a time-step $\delta t=1$~fs, employing the Nose-Hoover thermostat and barostat with damping times of $10^2$ and $10^3$ time steps, for temperature and pressure respectively. The systems were initialized from structures generated with CHARMM-GUI~\cite{jo2008charmm}. We considered a large number of independent configurations, by varying both both the salt concentration, $c$, and the silica nanoparticles volume fraction, $\Phi_{\text{SiO$_2$}}$, in an extended range of temperatures, $T$. These were all subjected to an annealing period of 2~ns  at $T=$~1000~K, followed by a 1~ns run where the temperature was decreased to the desired value. The systems were next further annealed, linearly raising the temperature to T=1000~K in 1~ns, followed by an additional 2~ns at this temperature. $T$ was finally lowered to the final value over 1~ns. We repeated the above annealing process twice, to eliminate any memory of the the initial configuration. The system was finally further equilibrated for 8~ns, followed by the 60~ns production runs, where we collected the systems configurations for the analysis. 
\subsection{Measured observables}
\label{subsect: observables}
We now introduce the structural and dynamical quantities that we will discuss below, expressed at the microscopic level in terms of the atomic system coordinates. The pair distribution function, $g(r)$, is a fundamental tool in the characterization of the structure of liquids, where the absence of long-range order makes traditional crystallographic methods impractical. Here we will focus on the partial $g_{\alpha\beta}(r)$, which quantifies the probability of finding a particle of type $\beta$ at a distance $r$ from a reference particle of type $\alpha$,
\begin{equation}
g_{\alpha\beta}(r)=\frac{V}{N_\alpha N_\beta}\sum_{i=1}^{N_\alpha}\sum_{j=1}^{N_\beta}\langle \delta(r-|\vec{r}_{j}-\vec{r}_i|)\rangle.
\label{RDF_eq}
\end{equation}
Here, $N_\alpha$ and $N_\beta$ are the number of atoms of type $\alpha$ and $\beta$, respectively, and $V$ is the volume of the simulation box. The $g_{\alpha\beta}(r)$ are determined by populating a histogram with the distances between all pairs of particles. Even though the $g(r)$ contains information on both short and long-range spatial correlations, we will focus on the information pertaining to distances that encompass the first main peak of these functions, corresponding to the first coordination shell of the reference atom. We are, in particular, interested in the coordination numbers,
\begin{equation}
C^{\beta}(\alpha)= \frac{4\pi N_\beta}{V} \int_{0}^{r_{m}} dr\, r^{2} g_{\alpha\beta}(r),
\label{eq:coordination}
\end{equation}
where $r_{m}$ is the distance at which the first minimum of $g_{\alpha\beta}(r)$ occurs. The $C^{\beta}(\alpha)$ count the number of nearest neighbors atoms of type $\beta$ around particles of type $\alpha$, and are an important measure of the local structure, providing valuable insight into the packing and bonding features of atoms in a material. This is particularly relevant in the present context, where the coordination environment of the mobile ions can strongly affect the ionic conductivity.

Our discussion of ionic transport in the investigated materials is mainly based on diffusion data. We have calculated the mean squared displacement of ion of type $\alpha=$~+ or -, defined as 
\begin{equation}
\langle r_\alpha^{2}(t) \rangle = \frac{1}{N_\alpha}\langle\sum_{i=1}^{N_\alpha}\left| \vec{r}_{i}(t)-\vec{r}_{\,i}(0)\right|^{2}\rangle,
\label{eq:msd}
\end{equation}
where $\vec{r}_{i}(t)$ is the position of atom $i$, at time $t$, and $\langle \rangle$ is an average over the MD trajectory. At short times, the mean squared displacement varies as $t^2$ (ballistic regime), followed by a crossover to the long-times linear(Fickian) regime~\cite{rehage1970fickian}, where one can extract the self-diffusion coefficient from the Einstein relation (in three dimensions),
\begin{equation}
D_\alpha = \lim_{t\to\infty}\frac{\langle r_\alpha^2(t) \rangle }{6t}.
\label{eq:diffusion}
\end{equation}
This is a crucial quantity that allows us to determine the ionic conductivity in the Nernst-Einstein approximation,
\begin{equation}
\sigma_{NE}=\sigma=\frac{N_{pair}}{V k_B T}(Q_+^2D_++Q_-^2D_-)=\sigma_+ + \sigma_-,
\label{eq:sigma_NE}
\end{equation}
where $N_{pair}$ is the number of ion pairs, $Q_+$ and $Q_-$ are the total charges on the cation and anion, and $D_+$ and $D_-$ are the diﬀusion coefficients of the cation and anion, respectively. $\sigma_+$ and $\sigma_-$ are the contributions to $\sigma$ associated to the cations and anions, respectively. Eq.~(\ref{eq:sigma_NE}) is only accurate in the limit of low ionic concentration, missing all contributions coming from charge fluctuations that could be relevant in our context. Based on many other works, however, we only expect quantitative modifications from much more complex calculations taking into account collective effects, which does not spoil the consistent qualitative picture based on Eq.~(\ref{eq:sigma_NE}). 

We have finally determined the cation transference number, which accounts for the contribution of lithium to the total ionic transport and is defined as,
\begin{equation}
t_+ = \frac{\sigma_+}{\sigma_++\sigma_-}.
\label{eq:transference}
\end{equation}
Note that $t_+$ is a critical parameter to be optimized in order to enhance the overall efficiency of the energy storage systems of interest here.
\begin{figure}[t]
\centering
\includegraphics[width=0.49\textwidth]{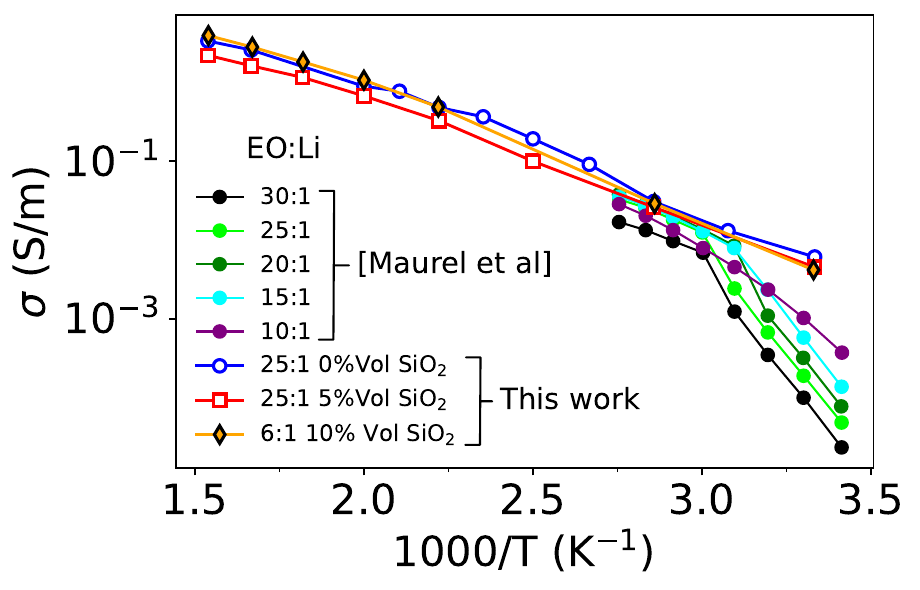}
\caption{Arrhenius plot of the ionic conductivity, $\sigma(T)$, as a function of the inverse temperature, for the indicated systems compositions. The MD data (open symbols) have been calculated from Eq.~(\ref{eq:sigma_NE}), both for the pure system and the hybrid electrolyte. The experimental data have been obtained by electrochemical impedance spectroscopy (EIS) in Ref.~\cite{maurel2020poly}. These data are discussed at length in the main text.}
\label{fig:Temp_dependence}
\end{figure}
\begin{figure}[t]
\centering
\includegraphics[width=0.48\textwidth]{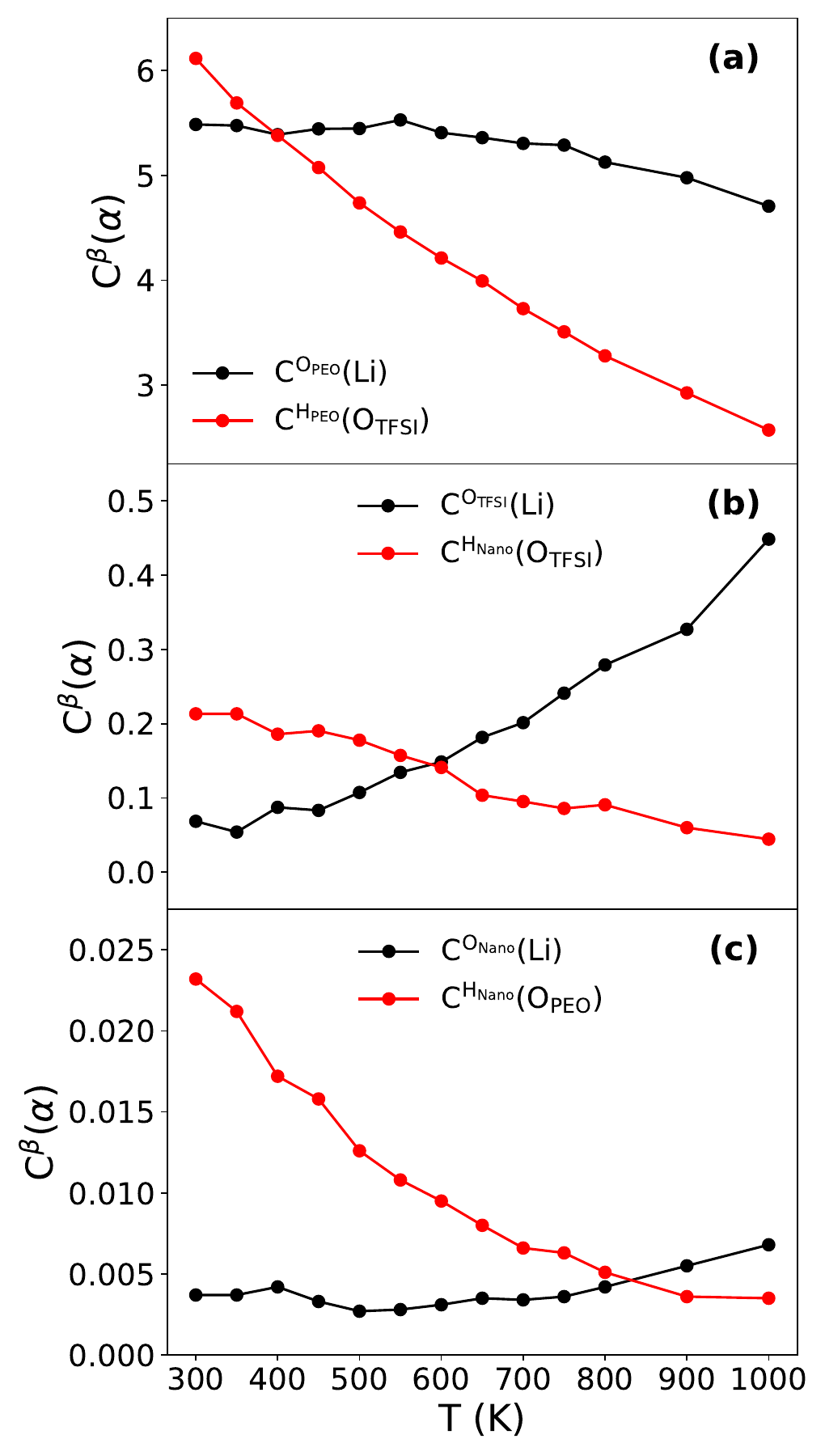}
\caption{$T$-dependence of the indicated coordination numbers, $C^\beta(\alpha)$, calculated from Eq.~(\ref{eq:coordination}). We recall that $C^\beta(\alpha)$ counts the number of nearest neighbors atoms of type $\beta$ around a particle of type $\alpha$. The ionic concentration was fixed to $\lambda^*=$~3.2, which corresponds to the maximum of conductivity, as discussed in the following. A 2\% in volume silica nanoparticles content was included in the system. These data are discussed in depth in the main text.}
\label{fig:CN-Temperature}
\end{figure}
\section{Results}
\label{sect:results}
\subsection{Effect of temperature and comparison to experiments}
\label{subsect: temperature dependence}
We start the discussion by testing our MD transport data versus available experimental measurements. Although we do not expect quantitative agreement between the two sets of data, due to our crude FF which disregards polarization effects among other possibly important mechanisms, previous studies are reassuring about the degree of realism of our modeling detail. In Fig.~\ref{fig:Temp_dependence} we show a comparison of the conductivity calculated from Eq.~(\ref{eq:sigma_NE}) (open symbols) with the experimental data of~\cite{maurel2020poly} (closed symbols, no nanoparticles) on completely amorphous PEO, at the the indicated TFSi concentrations. 

We note that above the melting temperature, $T_m\simeq 65 ^oC$ (the melting temperature of the salt in the polymer matrix can vary, but it is typically in the range of 60 to 80~$^o$C for PEO-based electrolytes with LiTFSI concentrations commonly used in battery applications), the numerical data are in good qualitative agreement with the experimental conductivities. In particular both sets of data follow an Arrhenius-like behavior, $\sigma(T)=\sigma_\infty\exp\{\Delta E/R\; T\}$, with very similar values of the activation energy $\Delta E \simeq 32.39$~J/mol, while the prefactor, $\sigma_\infty$, is slightly higher for the simulation data. This discrepancy is not surprising if one notes that the size of the PEO chains investigated in experiments ($>10^{6}$~g/mol) is much larger than the one of the model chain considered here ($\sim 10^{3}$~g/mol). This already explains the slight disparity in conductivity, since smaller chains exhibit higher mobilities and, indeed, the difference in diffusion corresponding to these two sizes is found to be about a factor of 2~\cite{devaux2012mechanism}. 

The simulation results follow the same Arrhenius behavior in the entire investigated $T$-range, even below $T_m$. This is at variance with the experimental data, where a clear crossover to an exponential behavior with a substantially higher $\Delta E$ appears. This is, again, not surprising, since MD simulations are limited in timescale and unable to accurately capture the freezing process mainly due to the entanglement of the polymer chains, yielding a liquid-like behavior even at temperatures below $T_m$. We observe the same temperature dependence when SiO$_2$ nanoparticles at the indicated concentrations are included in the simulations, although no equivalent experimental data are available in this case. All together, our model seems to reasonably catch the main features of the macroscopic ionic transport in the actual electrolyte, both above and close to $T_m$.  

The above conductivity behavior must correspond to substantial modifications of the coordination properties of both co-ions. In Fig.~\ref{fig:CN-Temperature} we show the T-dependence of the coordination numbers, $C^\beta(\alpha)$ (Eq.~(\ref{eq:coordination})), among lithium, oxygens in TFSI and PEO, and hydrogens of PEO and SiO$_{2}$. (See Tab.~S1 for reference.)
These atoms play a major role in the inter-molecular interactions, due to their privileged position in the macro-molecules they pertain to, or to their high partial charges values. The ionic concentration is fixed to $\lambda^*=$~3.2, which corresponds to the maximum of conductivity, as we will see below. A $\Phi_{SiO_2}=$~2\% in volume silica nanoparticles content was included in the system.
\begin{figure}[t]
\centering
\includegraphics[width=0.48\textwidth]{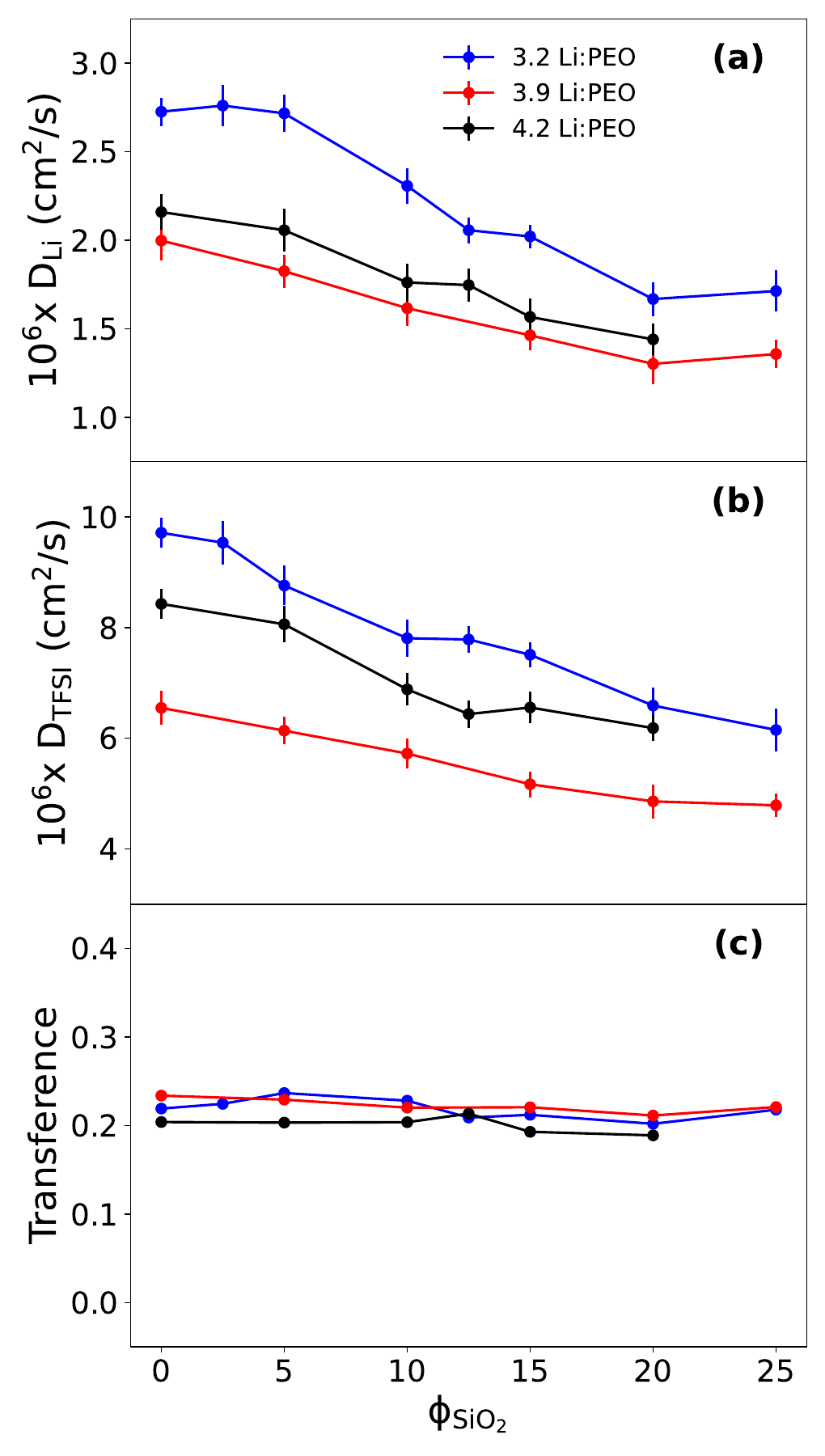}
\caption{\label{fig:DLiDAnTrans} Self-diffusion coefficient calculated from Eqs.~(\ref{eq:msd}) and~(\ref{eq:diffusion}) for Li$^+$~(a), and TFSI$^-$~(b), respectively, as a function of the SiO$_2$ nanoparticles volume fraction, at the indicated values of the ionic concentration, and at $T=$600~K. In~(c) we plot the transference number calculated from Eq.~(\ref{eq:transference}) as a function of the SiO$_2$ content.}
\end{figure}
\begin{figure}[t]
\centering
\includegraphics[width=0.48\textwidth]{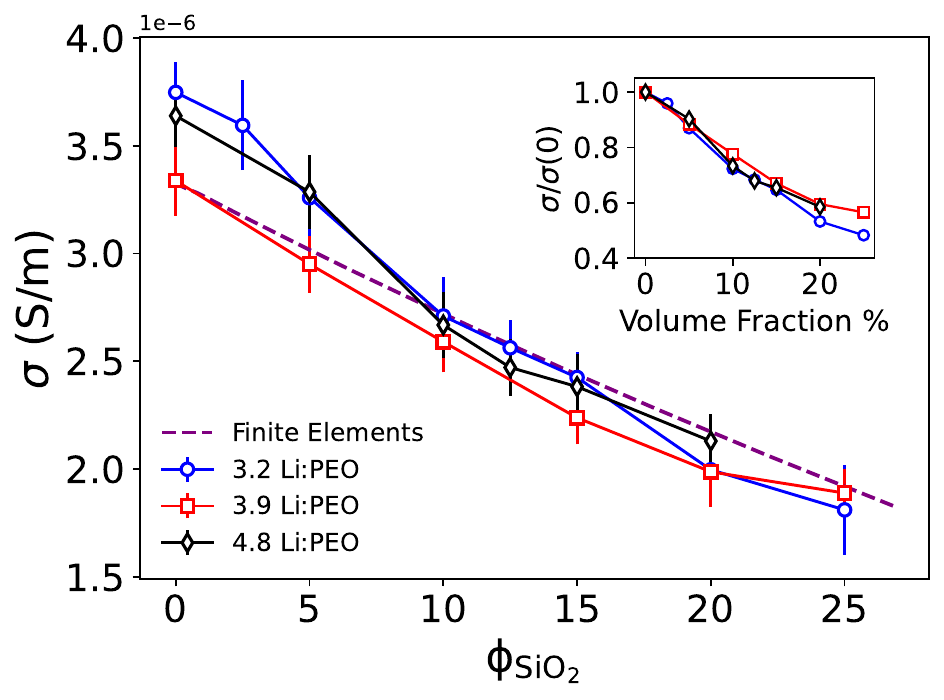}
\caption{Main panel: Conductivity as a function of volume fraction of SiO$_{2}$ at 600~K. The open symbols are the MD data, the dashed line corresponds to the finite elements approximation to the diffusion equation, discussed at length in the main text, for the concentration $\lambda=$4.17. Inset: same MD data as in the main panel, normalized as $\sigma(\lambda)/\sigma(\lambda=0)$. These data are discussed in depth in the main text.}
\label{fig:Nano_dependence}
\end{figure}

We start by noticing that all C$^\beta(\alpha)$ exhibit monotonous behaviors in the entire $T$-range, without the appearance of any significant distinct abrupt features, pointing toward continuous structural modifications with $T$. In Fig.~\ref{fig:CN-Temperature}(a) we can observe that, despite the presence of the silica nanoparticles, the lithium ion strongly coordinates, on average, with $\simeq 5.5$ O-atoms pertaining to the PEO chains at ambient temperature, a number which only slightly decreases to $\simeq 4.5$ at the highest temperatures. The situation for the anions is similar although even more pronounced, with an average number of surrounding $H_{PEO}\simeq$ 6 at ambient $T$ decreasing of more than a factor of 2 at the highest $T$. These numbers are in agreement with previous studies ~\cite{molinari2018effectKozinsky,diddens2010understanding,johansson1999modelling,deng2021role} of the pristine system, and point to the well known situation where the formation of ion pairs is inhibited by an overwhelmingly high coordination of both co-ions with the PEO. 

Interestingly, the addition of nanoparticles only has minor (although interesting) effects on the anion coordination. Indeed, at ambient temperature, a few TFSi are adsorbed at the surface of the nanoparticles (Fig.~\ref{fig:CN-Temperature}(b)), while the content of PEO and Li directly coordinating with silica is negligible in the entire $T$-range (Fig.~\ref{fig:CN-Temperature}(c)). By increasing $T$, those anions increasingly desorb from the solid surface (Fig.~\ref{fig:CN-Temperature}(b), red symbols) and progressively enter the co-ion coordination sphere (Fig.~\ref{fig:CN-Temperature}(b), black symbols) allowing the formation of a small amount of ion pairs. We note that, despite the above modifications, most part of the Li-ions remain dissolved in the polymer, resulting in a substantial increase with temperature in both atomic diffusion coefficients and material conductivity, due to the enhanced thermal energy. 

Overall these data seem to suggest that the modifications induced by the interaction of the ions with a fixed amount of nanoparticles active surface are small, although non negligible, and substantially overwhelmed by the variations induced by temperature. In the following we attempt to complete the general picture, by modulating both the nanoparticles content and the ions concentration (charge loading), by keeping $T$ constant. 
\subsection{Impact of SiO$_2$ nanoparticles volume fraction}
\label{subsect: impact nanoparticles}
In Fig.~\ref{fig:DLiDAnTrans} we show the impact of the SiO$_{2}$ nanoparticles volume fraction on the diffusion coefficient, Eq.~(\ref{eq:diffusion}), at the indicated ionic concentrations, $\lambda=$~3.2 , 3.9 and 4.2 at $T=$600~K, for Li and TFSi in Fig.~\ref{fig:DLiDAnTrans}(a) and (b), respectively. (The point at the origin consistently corresponds to the pristine solid electrolyte, without nanoparticles. We also note that $\lambda^*=$~3.2 corresponds to a concentration $c^*=$~2 mol/l.) The latter two values of $\lambda$ slightly exceed $\lambda^*$, allowing us to investigate changes at the transition to the saturated regime, that we define below. We first observe that the anion diffusion is consistently higher than that associated to the cation, a feature common to many electrolytes. In addition, in all cases we find that enhancing the silica content degrades the ionic diffusivity, for both ions. The transference numbers calculated from Eq.~(\ref{eq:transference}) (Fig.~\ref{fig:DLiDAnTrans}(c)), in contrast, keep constant values between 0.2 and 0.3, indicating that the relative contributions of the co-ions to transport are not significantly modified in HSEs compared to the SPE in the entire investigated $\Phi_{\text{SiO$_2$}}$-range.
\begin{figure}[b]
\centering
\includegraphics[width=0.48\textwidth]{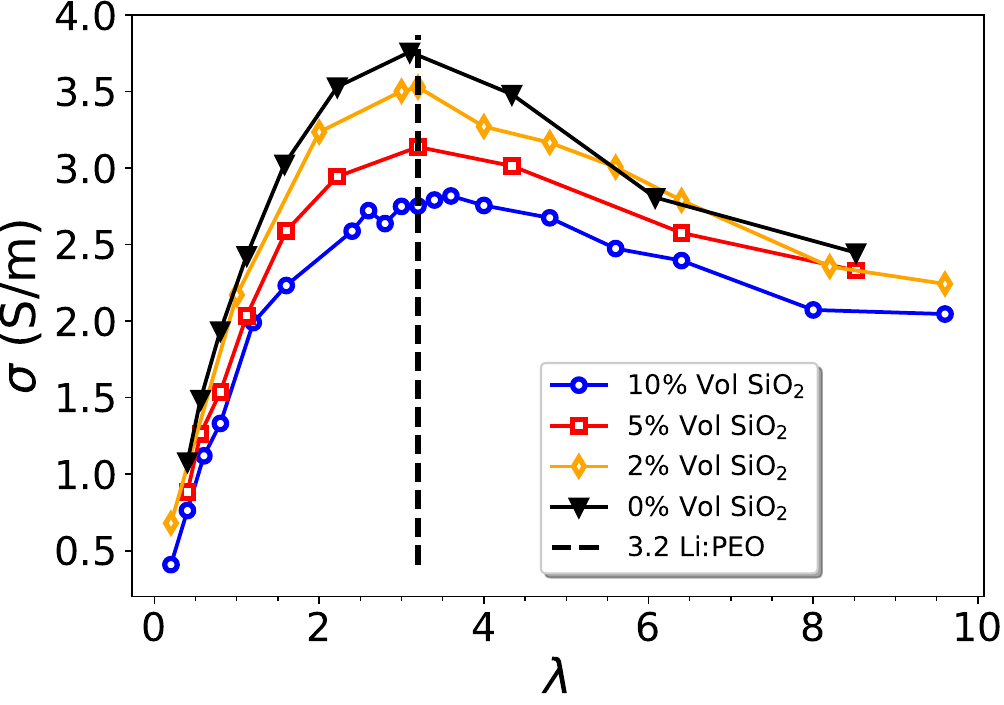}
\caption{Conductivity, $\sigma$, as a function of the ionic concentration at the investigated values of SiO$_{2}$ loading. $\sigma$ for the pristine case is also shown, for reference. The vertical dashed line indicates the optimal concentration $\lambda^*=$~3.2 ($c^*\simeq$~2 mol/l) corresponding to the maximum of $\sigma$. These data are discussed at length in the main text.}
\label{fig:IonicsConcentration}
\end{figure}

In Fig.~\ref{fig:Nano_dependence} we show the dependence of the electrolyte conductivity on the silica content, extracted by inserting the data of Fig.~\ref{fig:DLiDAnTrans} in Eq.~(\ref{eq:sigma_NE}). These data confirm that the addition of SiO$_2$ nanoparticles continuously degrades the ionic conductivity of the hybrid material, halving the value pertaining to the pristine material at the highest silica contents (see the normalized $\sigma(\Phi)/\sigma(0)$ in the inset). The origin of this behavior can be traced back to an apparently simple mechanism. Indeed, the dashed line in Fig.~\ref{fig:Nano_dependence} is a finite-elements solution of a diffusion equation for an ionic concentration of Li:PEO$=$4.2, where we treat the SiO$_{2}$ nanoparticles as a portion of space inaccessible to ions, in the shape of spherical cavities arranged as a simple cubic lattice inside the diffusive medium. (We have employed the NDSolve function in Mathematica 11.13~\cite{Mathematica}, to solve the Laplace equation in the "ImplicitRegion" between the four faces of a cube of side $L$, and the surface of a sphere of radius $r < L/2$ centered in the middle of the cube. We have used Dirichlet boundary conditions to ensure a zero gradient perpendicular to the sphere surface or to the four cube faces perpendicular to transport, and a finite gradient perpendicular to the other two cube surfaces.)
\begin{figure}[t]
\centering
\includegraphics[width=0.48\textwidth]{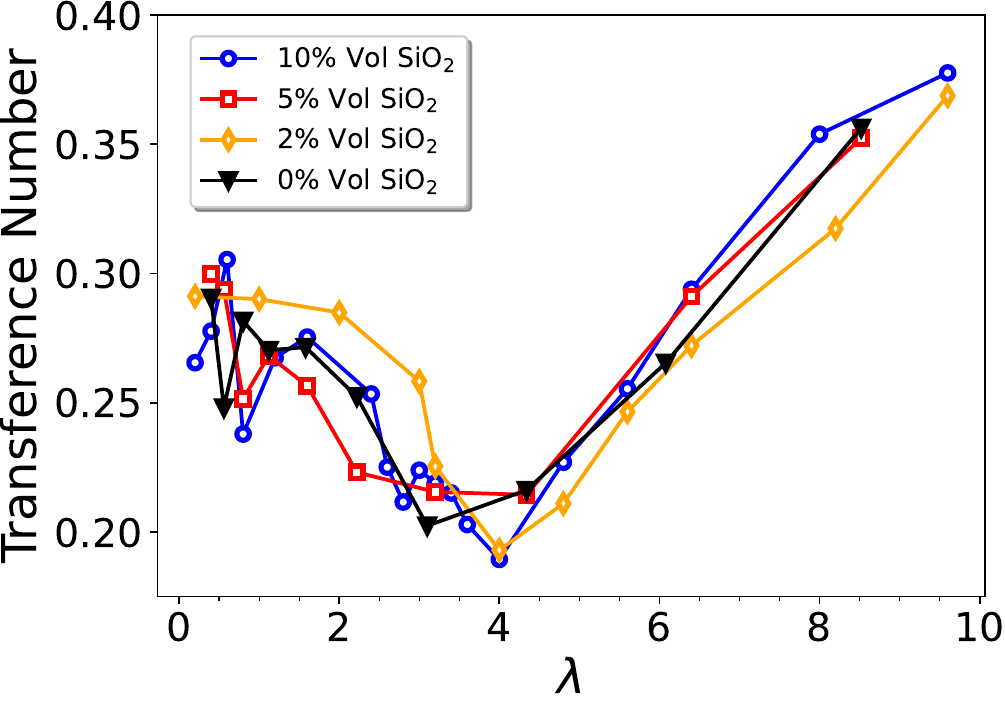}
\caption{\label{fig:IonicC_Trans} Transference number of lithium as a function of ionic concentration for the same systems than those in Fig.~\ref{fig:IonicsConcentration}.}
\end{figure}

Remarkably, this simple model exhibits a trend very close to our MD simulation data (open squares), although systematically slightly higher. Other analytical and numerical approximations for the conductivity of a suspension of empty spheres also provide similar results~\cite{pavlin_effective_2002}. Also, recent experimental measurements on well-dispersed polymer nanocomposites have reported reductions of $\sigma$ below the Maxwell or Bruggeman models \cite{maxwell1954electricity}, suggesting the existence of additional effects related to hindered segmental motion of the polymer in the proximity of the nanoparticles~\cite{tekell2023ionic}, which is consistent with the MD simulations of~\cite{hanson2013mechanisms}. 

In our simulations, the observed detrimental effect on the conductivity due to the addition of the nanoparticles hence seems to consist in a mere geometrical effect, due to the {\em excluded volume} represented by the nanoparticles themselves and unavailable to ionic diffusion. This is at variance with the competing possibility that transport could be influenced by the interaction of ions with the atoms localized at the available solid active {\em surface} \cite{croce2001role,wang2017lithium,hu2021composite}. We note at this point that the above findings are corroborated by recent work, including~\cite{isaac2022dense} that reports experimental data on several active ceramics, reaching the conclusion that the addition of nanoparticles does {\em not} enhance the conductivity of the polymeric electrolyte. Other studies on passive ceramics, such as~\cite{chen2019study,chen2018mechanical}, provide further evidences in the same direction. In \cite{chen2019study} , for instance, it was demonstrated that the incorporation of a 30\% volume fraction of ceramic nanoparticles resulted in a conductivity decrease of 60\% compared to the pristine value. And, indeed, this value is very close to what we can expect in our system for a 30\% volume content of SiO$_{2}$.
\begin{figure}[b]
\centering
\includegraphics[width=0.48\textwidth]{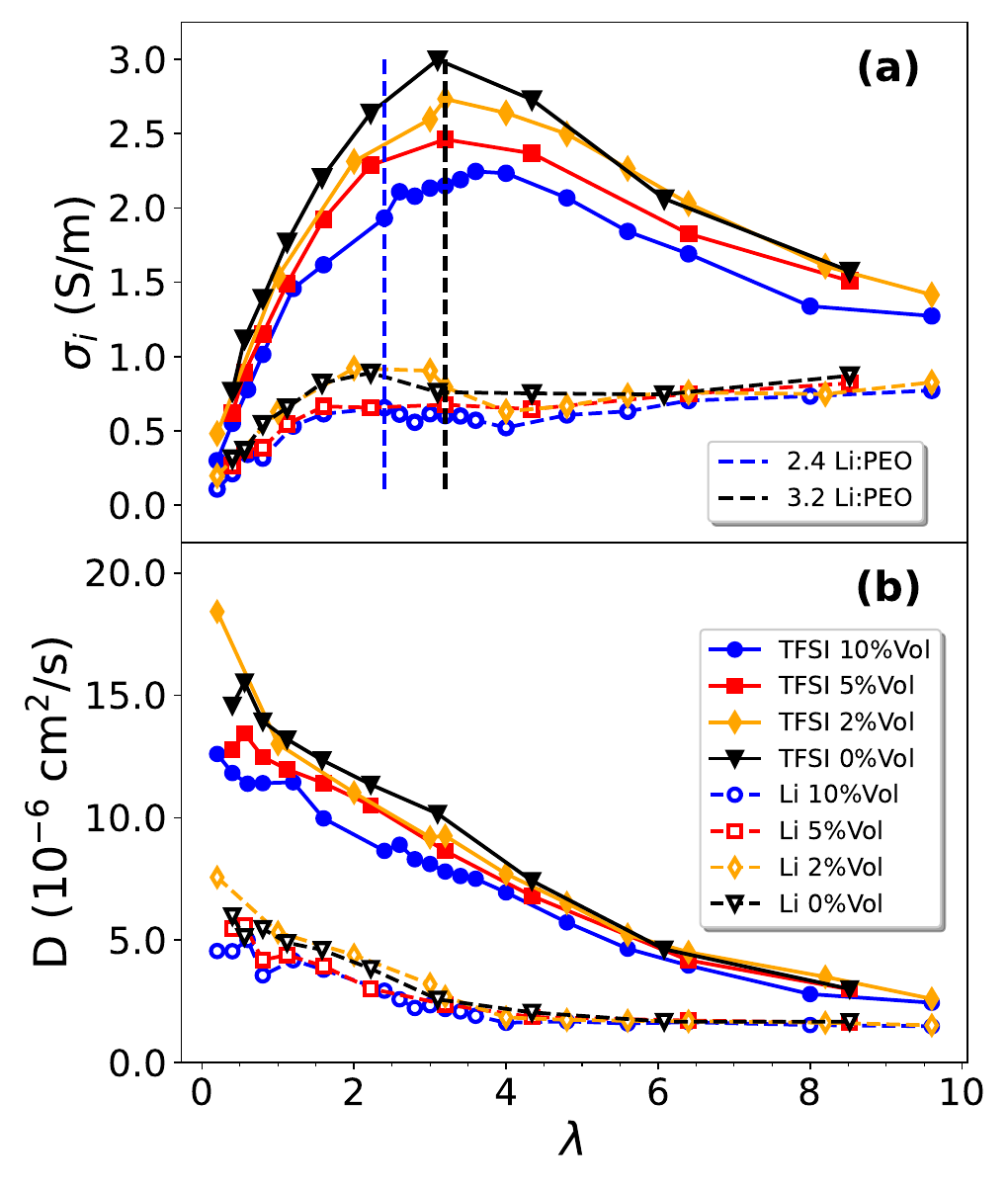}
\caption{Conductivity and diffusivity as a function of ionic concentration for lithium (solid lines) and TFSI (dashed lines).}
\label{fig:IonicCLi&TFSI}
\end{figure}
\begin{figure*}[t]
\centering
\includegraphics[width=0.7\textwidth]{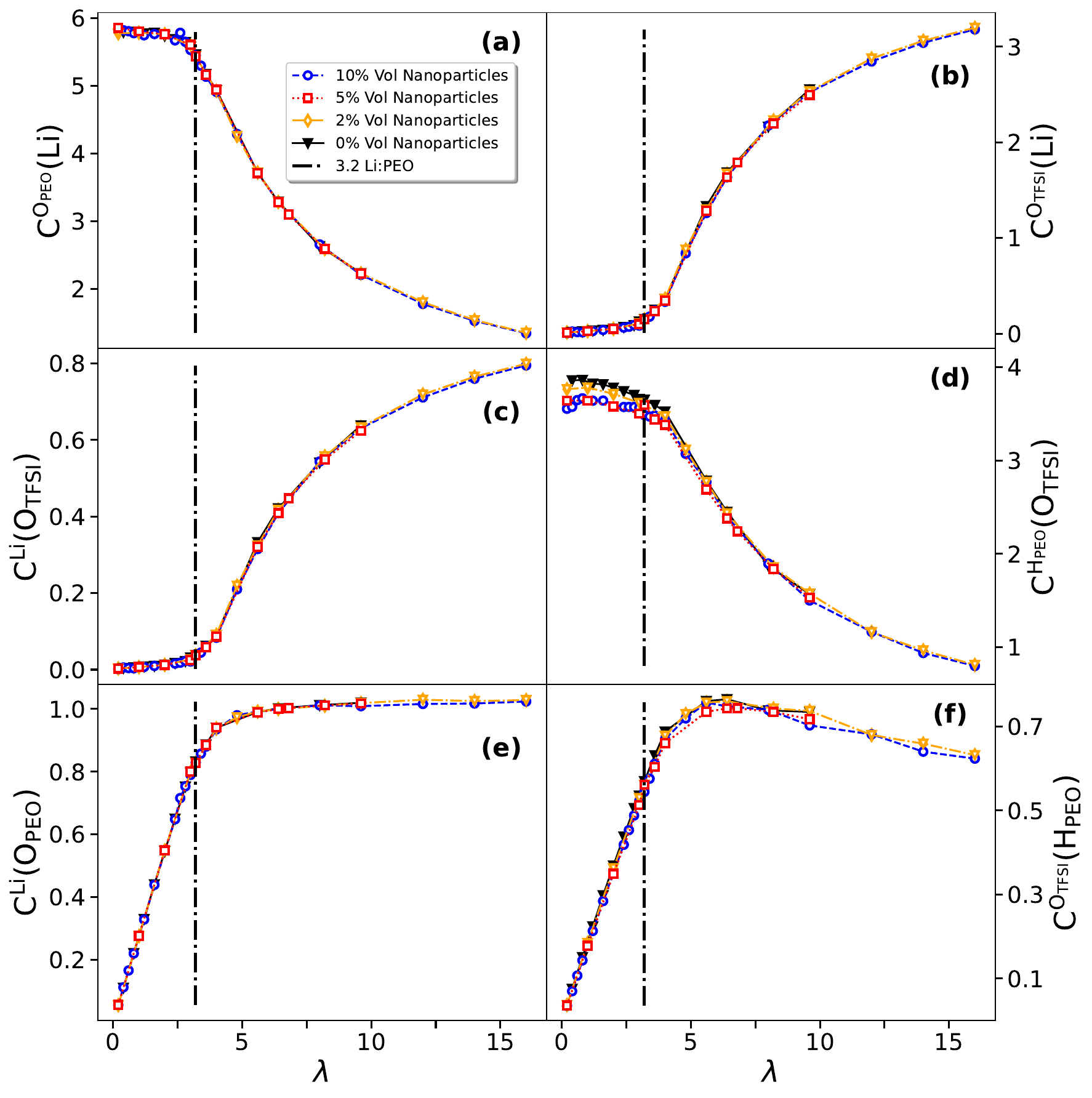}
\caption{\label{fig:VI-T600-N4-All} Coordination number between (a) lithium and oxygen of PEO, (b) lithium and oxygen of TFSI, (c) oxygen of TFSI and lithium, (d) oxygen of TFSI and hydrogen of PEO, (e) oxygen of PEO and lithium, (f) hydrogen of PEO and oxygen of TFSI. The four different colors correspond to the same loadings as in Fig.~\ref{fig:IonicsConcentration} (see inset).}
\end{figure*}
\subsection{Effect of ion concentration}
\label{ion concentration}
We improve the above picture by investigating the ion concentration impact on transport at fixed nanoparticles concentration, $\Phi_{\text{SiO$_2$}}$. In Fig.~\ref{fig:IonicsConcentration} we show the dependence of conductivity on ionic concentration, at the indicated values of the nanoparticles content and for the pristine material (black closed triangles). $\sigma$ conforms to what is generally expected for electrolytes in this range of concentrations, increasing at low $\lambda$ due to the addition of free ions (low concentration regime, LC), going through a maximum at $\lambda^*$ (equivalent to $c^*=$~2 mol/L, in agreement with~\cite{molinari2018effectKozinsky}), and eventually decreasing at high $\lambda$ (high concentration regime, HC), due to the formation of an increasingly high number of neutral ion pairs, which do not contribute to $\sigma$. The hybrid systems show a similar qualitative picture, keeping the same general behavior at all silica concentrations. Interestingly, while the optimal concentration is unchanged when gradually increasing the nanoparticle content up to 10\% in volume, $\sigma$ decreases consistently. Here, again, even in ionic loading conditions where $\sigma$ is optimal, the pristine system seems to be the most efficient choice when high conductivity is required.

Remarkably, we find a quite complex behavior of the transference numbers, that we show in Fig.~\ref{fig:IonicC_Trans}, at the indicated values of nanoparticles loading. $t_+$ assumes a constant value in the range $\lambda\simeq$ 0.5 to 1.50, at low $\lambda$, goes through a {\em minimum} at the optimal $\lambda^*$ where $\sigma$ is maximum, and keeps increasing at higher concentrations. This trend is similar to that obtained in a recent simulation work for bulk PEO~\cite{shao_transference_2022}, which itself confirmed earlier experimental results of~\cite{villaluenga_negative_2018}. (The quantitative differences of about 10\% of our data with respect to those results most probably stem from the neglect of the Onsager cross terms in the Nernst-Einstein approximation.)

Based on our calculations, we can be specific about the $t_+$ modifications going from the LC to the HC regimes through the optimal concentration $\lambda^*$. Indeed, the results of Fig.~\ref{fig:IonicC_Trans} can be understood by inspecting the individual contributions, $\sigma_+$ (open symbols) and $\sigma_-$ (closed symbols), shown in Fig.~\ref{fig:IonicCLi&TFSI}(a) at the indicated values of $\Phi_{\text{SiO$_2$}}$. From these data it is clear that the anions dominate the transport features, not only accounting for more than 80\% of $\sigma(\lambda^*)$, but also determining the overall $\lambda$-dependence, see Fig.~\ref{fig:IonicsConcentration}. Interestingly, the conductivity of lithium follows a different pattern, with a much less intense maximum at a lower ionic concentration $\Tilde{\lambda}\simeq$~2.4, followed by a shallow {\em minimum} around $\lambda^*$, for eventually reaching at $\lambda\simeq$ 6 a high-$\lambda$ value $\sigma_+\simeq$~0.8 S/m (corresponding to a limiting $t_+\simeq$~1/2 in Fig.~\ref{fig:IonicC_Trans}). This behavior is mirrored in the ionic concentration dependence of the diffusivity of the co-ions, Fig.~\ref{fig:IonicCLi&TFSI}(b), with a consistent decrease of both quantities going from the LC to the HC regime, while $\Tilde{\lambda}$ and $\lambda^*$ identify the position of inflection points for $D_+$ and $D_-$, respectively. 

The behavior of $t_+$ in Fig.~\ref{fig:IonicC_Trans} is therefore clear: for $\lambda<\Tilde{\lambda}$, both $\sigma_+$ and $\sigma_-$ increase at constant rates, keeping $t_+$ constant. At the intermediate $\Tilde{\lambda}<\lambda<\lambda^*$, in contrast,  $t_+$ decreases following the decrease of $\sigma_+$, reaches a minimum corresponding to the minimum of $\sigma_+$ (maximum of $\sigma_-$) at $\lambda^*$, and eventually continuously increases for $\lambda>\lambda^*$, due to an almost constant $\sigma_+$ and a strongly decreasing $\sigma_-$. In the following we correlate $\Tilde{\lambda}$ and $\lambda^*$ to the co-ions coordination features.
\begin{figure}[t]
\centering
\includegraphics[width=0.48\textwidth]{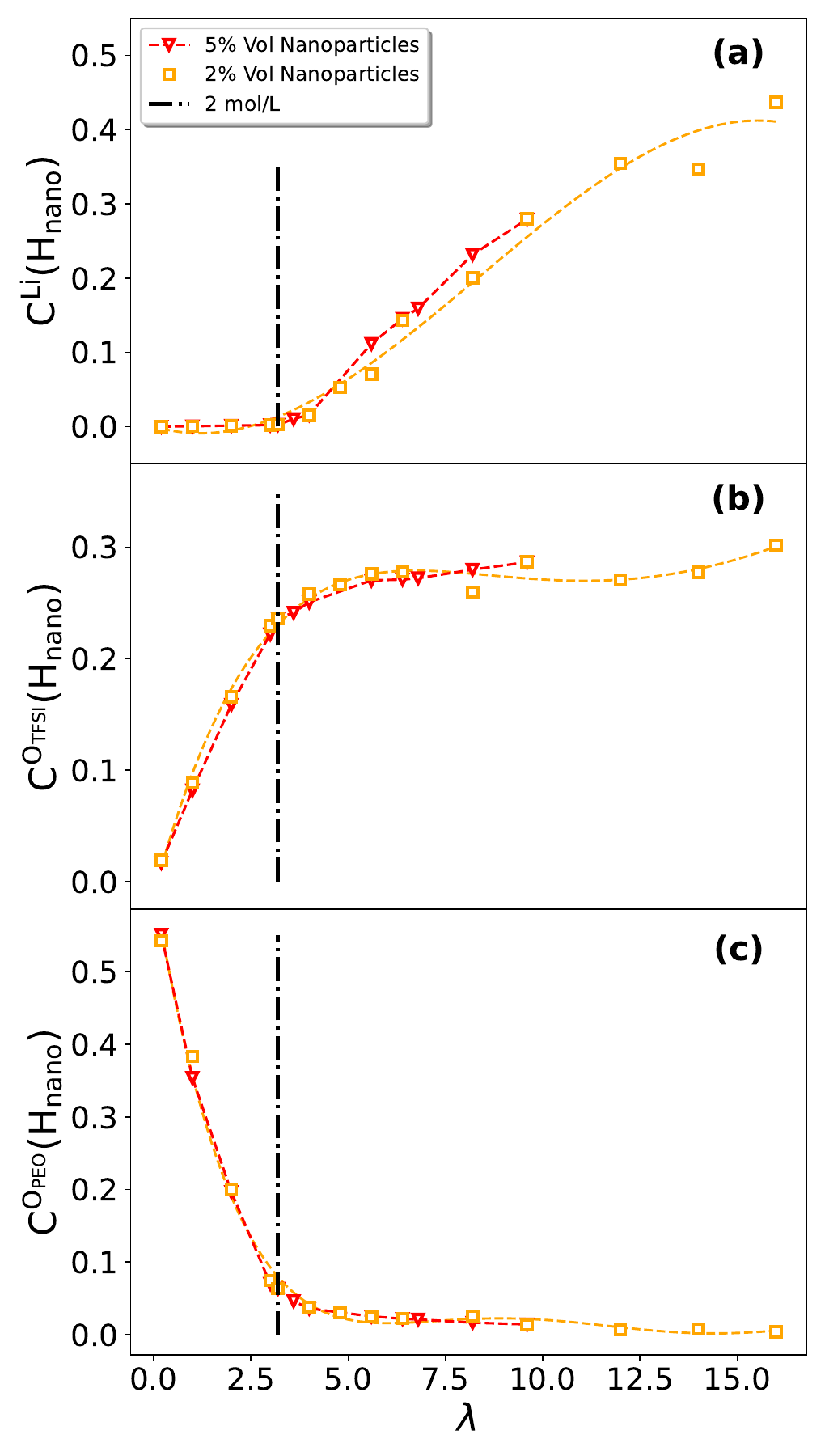}
\caption{\label{fig:VI-T600-N4-Si} Coordination number of some species in the coordination sphere of the surface hydrogens of the SiO$_2$ nanoparticles, for different concentrations of nanoparticles: (a) lithium, (b) oxygen of TFSI and (c)oxygen of PEO. Dashed lines are interpolations used as guides for the eyes. 
}
\end{figure}

In Fig.~\ref{fig:VI-T600-N4-All} we show the $\lambda$-dependence of the coordination numbers for the indicated chemical species. For $\lambda< \Tilde{\lambda}$, the coordination environment of each lithium ion is entirely composed by PEO, with $\simeq$ 6~$O_\text{PEO}$ folding around each cation (Fig.~\ref{fig:VI-T600-N4-All}~a)). Therefore, each added $Li^+$ sequentially traps a (sterically) maximum number of $O_\text{PEO}$ (as also clear from the increasing but low value of $C^\text{Li}(O_\text{PEO})$ in Fig.~\ref{fig:VI-T600-N4-All}~e)). TFSi behaves similarly (Figs.~\ref{fig:VI-T600-N4-All}~d) and~ f)), with the only difference that now a slight dependence on $\Phi_\text{SiO$_2$}$ is visible. Each polymer strand is therefore increasingly occupied by alternating co-ions which are completely screened by the PEO and, therefore, have no possibility to mutually interact, as it is clear from  Figs.~\ref{fig:VI-T600-N4-All}~b) and~c).

Interestingly, $\Tilde{\lambda}$ identifies a sudden change in the coordination mechanism for $Li^+$, with the $O_\text{PEO}$ cages unfolding and $C^\text{Li}(O_\text{PEO})$ starting to decrease with $\lambda$. Note that a similar modification occurs for TFSi only at a quite higher concentration, making $\Tilde{\lambda}$ relevant for the behavior of the cation only, consistently with Fig.~\ref{fig:IonicCLi&TFSI}. No additional notable changes occur for $\Tilde{\lambda}<\lambda<\lambda^*$.

$\lambda^*$, in contrast, marks two changes for both co-ions.
First, it identifies the salt concentration sufficient to trigger the formation of ion pairs in the HC regime (Figs.~\ref{fig:VI-T600-N4-All}~b) and~c)), consistent with the incipient decrease of $\sigma$. Indeed, both co-ions are progressively released from the PEO strands on increasing $\lambda$, with a $C^\text{Li}(O_\text{PEO})$ saturating at $1$ at $\lambda\simeq$~5, while $C^\text{TFSi}(H_\text{PEO})$ goes through a maximum at the same concentration, for slightly decreasing at higher $\lambda$. These released charges are now able to interact strongly, and at the highest investigated concentration almost all co-ions participate in ion pairs. (We recall that the coordination number is calculated at the atomic level, not between the molecular centers of mass.) Note that we cannot exclude the formation of larger ionic aggregates, which we would expect to significantly depend on $\Phi_\text{SiO$_2$}$. Those features, unfortunately, cannot be established on the basis of $\Phi_\text{SiO$_2$}$-independent, very local observables like the $C^\beta(\alpha)$.

Second, $\lambda^*$ is associated to interesting modifications in the interaction of both PEO and ions with the silica nanoparticles, inducing significant modifications of the resulting structure of the interface. This is clear by inspection of the coordination numbers of the indicated moieties, that we show in Fig.~\ref{fig:VI-T600-N4-Si}. For $\lambda<\lambda^*$, no lithium ion approaches to the nanoparticle, consistently with the above observation that, at low charge content, their entire coordination sphere is crowded by atoms pertaining to the PEO. In the same range the latter, which is obviously strongly adsorbed on the SiO$_2$ at vanishing salt concentration, rapidly desorbs from the interface, consistently followed by the opposite behavior of the anions. For $\lambda>\lambda^*$, in contrast, the (direct) interaction of PEO with the nanoparticles is almost negligible, while the H$_\text{nano}$ atoms are on average surrounded by a number of both ions which is as high as $\simeq 0.3$ at the highest concentration. Note that, interestingly, these curves do {\em not} depend on the nanoparticles volume fraction at any investigated value of $\lambda$, signaling a substantial invariance of the features of the interface with the extent of the available active surface.  
\section{Discussion}
\label{sect:discussion}
The results described above significantly depart from previous studies which suggested that nanoparticle incorporation improves conductivity above the PEO melting temperature~\cite{zaman2019visualizing, croce1998nanocomposite,scrosati2000impedance}. Indeed, in contrast with those findings, we find that SiO$_{2}$ nanoparticles, at any volume fraction, actually {\em reduce} the conductivity of LiTFSI ions in PEO. Consequently, the formation of a complex interface at the boundary between the ceramic and the bulk polymer does {\em not} have any positive impact on the conductivity. This conclusion agrees with the experiments of~\cite{tekell2023ionic}, where significant differences between the earliest experiments on ceramic HSEs (published more than 20 years ago) and the much more recent reported data have been highlighted. 

Indeed, it exists a considerable body of experimental work conducted over the past two decades, including prominent studies by Croce {\em et al.}~\cite{croce1998nanocomposite,croce1999physical,croce2001role}, and Scrosati {\em et al.}~\cite{scrosati2000impedance,scrosati2001new}, that initially supported the notion that the inclusion of ceramic nanoparticles significantly enhances the electrolyte ionic conductivity. A lively debate addressed the question of how the chemical nature of the surface of passive ceramic fillers, including Al$_{2}$O$_{3}$ and SiO$_{2}$, influences the conductivity of the electrolyte. We note that the dispute also extends to active fillers, as their influence in bulk SPEs cannot be explained solely by the bulk properties of the ceramics but rather as an enhancement arising from the particle-melt interaction, as described in \cite{zaman2019visualizing}. Notably, in~\cite{croce2001role} it was reported that the degree of acidity of the nanoparticles surface plays an important role, with acidic surfaces inducing an important enhancement of conductivity performances, followed by neutral surfaces, and mild effects only originating from basic surfaces. (In this particular case, the experiment was based on Al$_{2}$O$_{3}$, but the same conclusions can be transferred to the case of SiO$_2$). This positive trend in experimental findings persisted for quite some time, and was further substantiated by various other works \cite{krawiec1995polymer,jayathilaka2002effect,dissanayake2003effect,croce1998nanocomposite,chung2001enhancement,scrosati2000impedance,croce1999physical,capuano1991composite,appetecchi2000transport,croce2001role,scrosati2001new,wang2017lithium,hu2021composite}. Correspondingly much fewer computational studies, including~\cite{wang2022lithium}, corroborated these findings. 

A more recent wave of research, however, particularly in computational studies, has introduced conflicting outcomes. These computational investigations have consistently demonstrated a detrimental impact on conductivity due to nanoparticle addition, thereby challenging the conventional understanding of nanoparticle effects in HSEs. For instance, ~\cite{scrosati2000impedance} has been contradicted by more recent computational studies~\cite{mogurampelly2015effect,mogurampelly2016influence,mogurampelly2016influence2} that have demonstrated a negative impact of the presence of fillers, with the most nefarious effect observed exactly in the acidic case, in complete contradiction with the conclusions of that work. Another dimension of complexity arises from Fullerton {\em et al.} \cite{fullerton2011influence}, which emphasized the role of humidity-dependent water uptake on conductivity. Their experiments revealed that the crystallinity of polymer electrolytes evolves differently under dry and humid conditions, ultimately impacting conductivity. This complexity underscores the critical role of humidity conditions in HSE behavior and raises questions about the reproducibility and reliability of experimental results in various environmental settings. 

We conclude by noticing that even the extremely limited number of simulation works that have reported a positive impact on ionic transport resulting from the inclusion of ceramic nanoparticles, ultimately display some discrepancies when compared to experimental results of the same sign. For instance,~\cite{wang2022lithium} demonstrated a conductivity enhancement of over 50\% at room temperature upon the addition of Al$_{2}$O$_{3}$ nanoparticles. However, the reported structural analysis surprisingly revealed that the salt tends to form a significant proportion of ionic pairs already at low ionic concentrations. Consequently, the conductivity improvement upon nanoparticles inclusion is attributed to their solvation effect on the salt. In our case solvation already occurs in the pure solid polymer electrolyte and, as a consequence, the addition of the SiO$_{2}$ nanoparticles does not further improve this property.
\section{Conclusions}
\label{sect:conclusion}
In conclusion, our analysis of the structural and dynamical properties of the studied systems indicates that SiO$_{2}$ is not a suitable candidate for enhancing the conductivity of PEO. The main reason for this is that SiO$_{2}$ shows limited interaction with the elements in the simulation. Additionally, while SiO$_{2}$ does not significantly affect lithium's mobility, its presence can cause a slight adsorption of ions on its surface, leading to a negative impact on the mobility of the counter-ions. As a result, the overall conductivity decreases. Our calculations demonstrate that the addition of silica to PEO interferes with the dynamics of the ions, occupying regions available for transport in the pristine electrolyte and, consequently,  decreasing ion mobility, contrary to earlier expectations.

Furthermore, our findings indicate that SiO$_{2}$ fails to improve conductivity under optimal conditions, i.e., when the polymer is entirely amorphous and at the optimal concentration. We cannot, however, exclude the hypothesis that the addition of SiO$_2$ may enhance the ionic conductivity when the polymer exhibits a crystalline phase, by reducing the polymer's crystallinity.

In summary, we have shed light on the limitations of SiO$_{2}$ as an enhancer for PEO conductivity, highlighting the importance of the polymer's state and concentration, in nanocomposite electrolytes. While SiO$_{2}$ may not prove beneficial in some contexts, it remains a valuable area of investigation for future advancements in hybrid electrolyte research. Further research may reveal how to leverage nanocomposite materials more effectively to optimize the performance of advanced battery technologies.
\begin{acknowledgments}
This work has been supported by the French National Research Agency under the France 2030 program (Grant ANR-22-PEBA-0002). S. M. also acknowledges support by the project MoveYourIon (ANR-19-CE06/0025) funded by the French "Agence Nationale de la Recherche".
\end{acknowledgments}
\bibliography{References-Manuscript}
\newpage
\section*{TOC Graphic}
\begin{figure}[]
\centering
\includegraphics[width=0.48\textwidth]{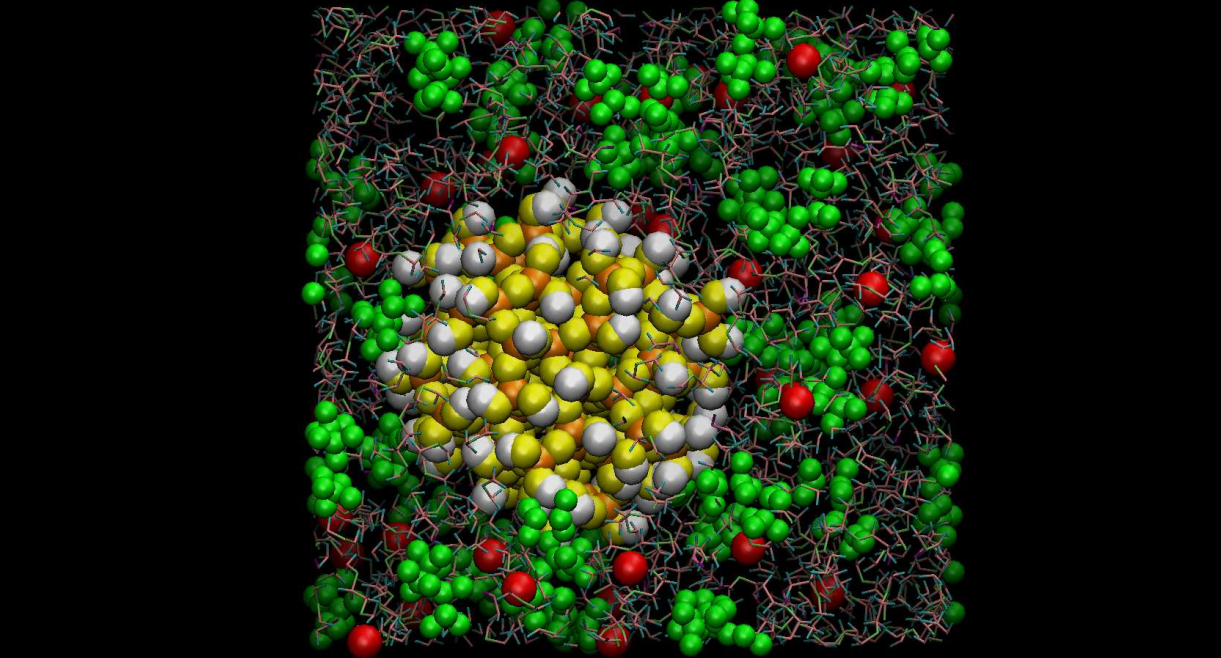}
\end{figure}
\end{document}